# Magnetic proximity effect at interface between a cuprate superconductor and an oxide spin valve

G.A. Ovsyannikov [a,b]*, V.V. Demidov [a], Yu.N. Khaydukov [c,d], L. Mustafa [c],

K.Y. Constantinian [a], A. Kalabukhov [b,d], D. Winkler [b]

[a] Kotel'nikov Institute of Radio Engineering and Electronics Russian Academy of Sciences, 125009, Moscow, Russia
[b] Chalmers University of Technology, SE-41296, Gothenburg, Sweden
[c] Max-Planck Institute for Solid State Research, D-70569, Stuttgart, Germany
[d] Institute of Nuclear Physics, Lomonosov Moscow State University, 119991, Moscow Russia

* gena@hitech.cplire.ru

Heterostructures consisting of a cuprate superconductor $YBa_2Cu_3O_{7-\delta}$ and a ruthenate/manganite ($SrRuO_3/La_{0.7}Sr_{0.3}MnO_3$) spin valve have been studied by SQUID magnetometry, ferromagnetic resonances and neutron reflectometry. It was shown that due to the influence of magnetic proximity effect a magnetic moment is induced in the superconducting part of heterostructure and at the same time the magnetic moment is suppressed in the ferromagnetic spin valve. The experimental value of magnetization induced in the superconductor has the same order of magnitude with the calculations based on the induced magnetic moment of Cu atoms due to orbital reconstruction at the superconductor-ferromagnetic interface. It corresponds also to the model that takes into account the change in the density of states at a distance of order of the coherence length in the superconductor. The experimentally obtained characteristic length of penetration of the magnetic moment into superconductor exceeds the coherence length for cuprate superconductor. This fact points on the dominance of the mechanism of the induced magnetic moment of Cu atoms due to orbital reconstruction.

1. Introduction

The penetration of ferromagnetic correlations into a superconductor (S) and superconducting correlations into a ferromagnet (F) attracts increased interest in recent years [1, 2]. In the case of a contact between ferromagnetic and normal metal the ferromagnetic correlations penetrate into the normal metal on a small (order of atomic size) distance near interface due to exchange interaction (see. e.g., [3]). It has been shown theoretically [4-6] that at S/F interface under the influence of exchange field a change in the density of states take place due to the differences for electrons with spin up and spin down. Later it was shown that the sign and magnitude of magnetic moment arising in the superconductor were highly dependent on the parameters of S/F interface such as transparency, presence of impurities and the thicknesses of layers [7-9]. Influence of parameters of the spin active barrier at S/F interface on the magnetic moment in the superconductor was considered theoretically in [10], and occurrence of an anomalous magnetic response in a normal metal in contact with the spin valve $S/F_1F_2$, generating triplet superconducting correlation are given in [11]. Experimental study of magnetic proximity effect (MPE) at S/F interface based on ferromagnetic metal and conventional superconductors were conducted by a variety of methods (ferromagnetic resonance, muon scattering, neutron scattering, etc.) [12-15] and, in general, confirmed the conclusions of the theory.

Among the structures made of cuprate superconductors with a small coherence length and anisotropic superconducting gap, the $YBa_2Cu_3O_7/La_{2/3}Ca_{1/3}MnO_3$ (YBCO/LCMO) superlattice (SL) was mainly studied. They revealed the presence of a magnetic moment in superconductor [16-20]. At the interface of cuprate superconductor and the magnetic material an induced magnetic moment of the Cu atom, oriented antiparallel relative to Mn atoms, was detected using X-ray dichroism [18-20]. It has been shown that the Cu and Mn atoms are connected through the interface by a covalent chemical bond, resulting in a strong hybridization and orbital reconstruction. Note that the characteristic length of orbital reconstruction greatly exceeds the interatomic distances and are 8-10 nm [21, 22]. Characteristic differences between the nuclear and the magnetic scattering profiles allowed to identify that a magnetic moment is induced in YBCO that is oriented antiparallel to the one in LCMO. On other side a magnetically "dead layer" arise in the LCMO region. [16,17]

The appearance of the magnetic moment in the cuprate superconductor contacting the manganite was considered in [22] theoretically. It was shown that as a result of the antiferromagnetic interaction of the spins of $x_2$-$y_2$ electrons of Cu and $e_g$ Mn a negative electron spin polarization is induced in the cuprate superconductor. The impact of this process on the properties of the superconductor is considerably stronger than the injection of spin-polarized electrons in ferromagnetic. An important parameter for the analysis of the processes is the depth of penetration of the magnetic moment of the superconductor, which is not limited to the adjacent layers of the interface but determined by the 8-10 atomic layers of a superconductor [20, 22].

The comparison of LCMO/YBCO и LMO/YBCO shows that the magnetic proximity effect is stronger at the interface of superconductors with ferromagnetic material which has better conductivity. Even the comparable magnetization of LCMO и LMO films the magnetic moment induced in YBCO films is one order larger in LCMO/YBCO then in LMO/YBCO SL [17]. Thus MPE is governed by the electronic (orbital) state of the FM manganite layers. Preliminary studies of superlattices consisting of cuprate superconductors (YBCO) and rutenate ferromagnets (SrRuO3) showed a significant difference from superlattices with manganite (LCMO) in the dependence of the magnetization on the temperature [23], but an increase in the induced magnetization in YBCO films because of the better (almost an order of magnitude) conduction of SrRuO3 were not observed

This paper presents the experimental study of the magnetic moment in a heterostructure containing a cuprate superconductor and a ferromagnetic spin valve, which is formed by two ferromagnetic layers. The interface between YBCO and ferromagnetic was made by SRO film. The magnetic moment measurements of the heterostructures were carried out by a SQUID magnetometer, neutron reflectometry and ferromagnetic resonance (FMR) technique. Analysis of the data obtained using these three methods allows to determine induced magnetic moment of the superconductor and the change in the magnetic moment in the ferromagnetic spin valve. The changes in magnetic moments of the individual layers in the heterostructures under the influence of triplet superconducting correlations in ferromagnetics, which are induced due to the non-collinear orientation of the magnetization vectors in the ferromagnetic films of the spin valve are discussed.

2. Experimental procedure and samples.

We investigated epitaxially grown thin-film heterostructures consisting of the $YBa_2Cu_3O_{7-\delta}$ (YBCO) superconductor and two ferromagnetic layers: ruthenate $SrRuO_3$ (SRO) and manganite $La_{0.7}Sr_{0.3}MnO_3$ (LSMO). The heterostructures were fabricated by laser ablation at a temperature of 700-800 °C and an oxygen pressure of 0.3-0.6 mbar. The thickness of the superconductor is in the range of 80-200 nm and the thickness of the ferromagnetic layers ranged from 5 to 20 nm (see. Table 1). The heterostructures were covered by a thin (20 nm) layer of gold. The substrates of 5x5 mm² (110)$NdGaO_3$ (NGO), (001)$LaAlO_3$ (LAO) and $(LaAlO_3)_{0.3}+(Sr_2AlTaO_6)_{0.7}$ (LSAT) were used. The magnetization vector of the epitaxial film LSMO deposited on the (110) NGO substrate or film (001) YBCO is placed in the plane of the substrate [24, 25], whereas the SRO film magnetization vector lies outside the plane for used substrates [26].

The detailed studies of the field and temperature dependences of the magnetization of the films and heterostructures were conducted using a SQUID magnetometer (MPMS 3, Quantum Design) in the VSM mode [27]. The plane of the substrate was set to the direction of the magnetic field within 1-2 degrees.

The results of measurements of the field dependence of the magnetic moment of the heterostructure Au/LSMO/SRO/YBCO (#1 in Table 1) are shown in Fig.1. The measurements were carried out at temperature $T = 100$ K which is higher than the critical temperature of the superconductor. The dependence of the magnetic moment for external magnetic fields $H$ oriented parallel and perpendicular to the substrate was measured. For the magnetic field parallel to the plane of the substrate the main contribution to the field dependence of the magnetic moment $m_{\|}(H)$ is given by the LSMO film, as for perpendicular field the SRO film gives the main contribution to $m_{\perp}(H)$. On the $m_{\|}(H)$ dependence, no influence of SRO film for $H<600$ Oe was observed (Fig. 1a). For the $m_{\perp}(H)$ dependence the hysteresis loop of the heterostructure is determined by SRO film. (Fig. 1.b)

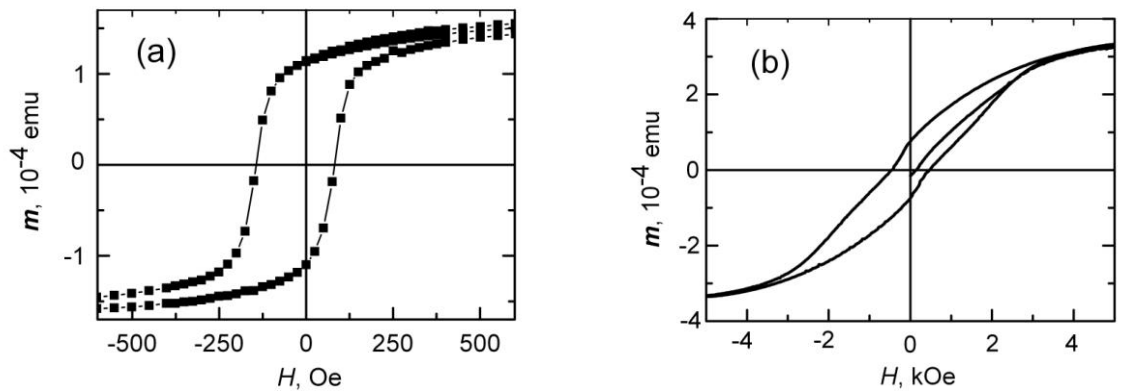

Fig.1. The hysteresis loop for heterostructure #1 Au/LSMO/SRO/YBCO at $T = 100$ K for the magnetic field: (a) oriented parallel and (b) perpendicular to the substrate plane.

The magnetic saturation field for the ferromagnetic films differs by more than an order of magnitude: from 0.1-0.3 kOe for LSMO films to 5-7 kOe for SRO films. Note that the parameters of the magnetic films in the heterostructure may be different from the parameters of films deposited directly on the substrate. In particular, the saturation field of the ferromagnetic films in the heterostructure is slightly smaller than that for films deposited directly on the substrate [25-26].

From the hysteresis loop the magnetization of the LSMO films at $T = 80$ K is equal to 2.5 $\mu_B$/Mn while for the SRO film the magnetization is equal 1.5$\mu_B$/Ru. The position of the easy axis of magnetization SRO film is close to the normal to the substrate plane. Noncollinear magnetization vectors of ferromagnetic films contributes to generation superconducting triplet correlations with non-zero spin projection of superconducting carriers in the ferromagnetic interlayer [28, 29].

The heterostructures were also investigated using a magnetic resonance spectrometer (Bruker ER 200), operating at the frequency 9.7 GHz. We measured the spectra of ferromagnetic resonance in a wide range of temperatures from 20 K to 300 K. The FMR spectra of the LSMO films in the heterostructures were obtained by cooling the sample in the field of the Earth. Upon reaching the desired temperature, the magnetic field was scanned from 0 to 4 kOe. The FMR spectrum of the SRO film in our experimental conditions can not be seen because of large values of the magnetic anisotropy field. The magnetic component of the microwave field $h_{MW}$ was directed perpendicular to the plane of the substrate. An external magnetic field $H_0$ is always lying in the plane of the substrate (see inset Fig. 2).

The angular dependence of the FMR spectrum of a thin ferromagnetic film in the presence of uniaxial and biaxial anisotropy for a case $H_0$ is lying in the plane of the film is described by the formula [30]:

$$\left(\frac{\omega}{\gamma}\right)^2 = \left(H_0 + H_u \cos 2\varphi_u + H_c \cos 4\varphi_c\right)\left(4\pi M_0 + H_0 + H_u \cos^2 \varphi_u + H_c \frac{1+\cos^2 2\varphi_c}{2}\right) \quad (1)$$

where $\gamma$ is gyromagnetic ratio, $H_u = 2K_u/M_0$, $H_c = 2K_c/M_0$, $K_u$ is the uniaxial anisotropy constant, $K_c$ is the constant of cubic anisotropy parameter, $M_0$ is the equilibrium magnetization in the absence of additional ferromagnetic layers, $\varphi_u$ and $\varphi_c$ are the angles for easy axis of the uniaxial and cubic anisotropy with respect to the external magnetic field, respectively. Note both the easy axis of uniaxial anisotropy $n_U$ and the easy axis of cubic anisotropy lie in the plane of the heterosructure. The following parameters of the LSMO ferromagnetic film in the heterostructure #2 at room temparature can be determined from the fitting of experimental data using the formula (1): $K_u = (8400\pm71)$ erg/cm$^3$, $K_c = (2250\pm66)$ erg/cm$^3$, $M_0 = (1.615\pm0.001)$ $\mu_B$/Mn, as well as the direction of the easy axes of both uniaxial and cubic anisotropy (Fig.2).

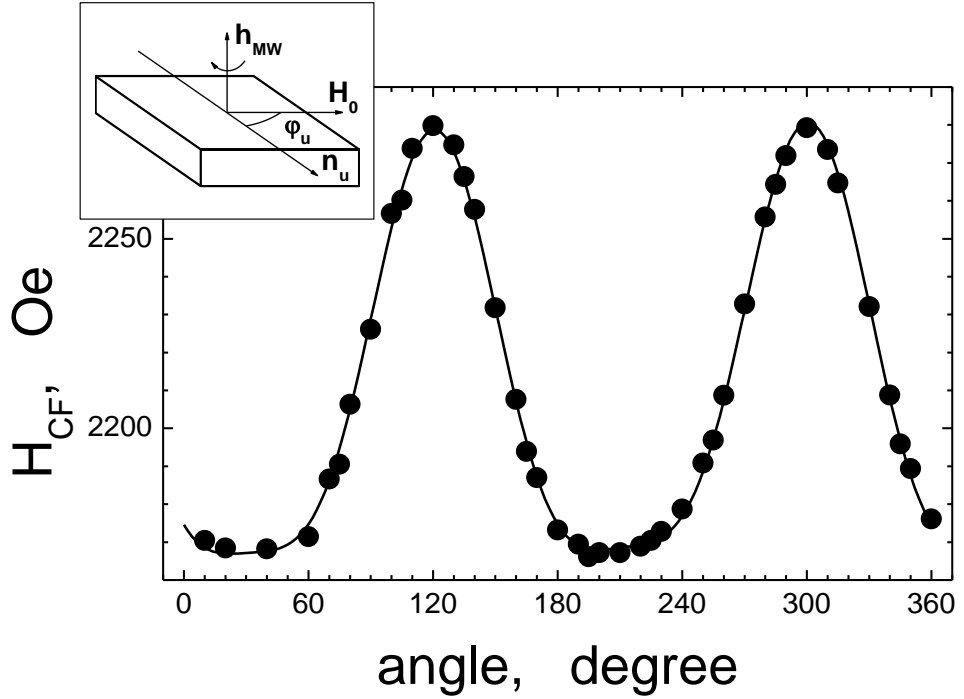

Fig. 2. The angular dependence of the resonant magnetic field at $T$=295 K for heterostructure #2. The solid line shows the relation (1) with fitted parameters: $K_u = (8400\pm71)$ erg/cm$^3$, $K_c = (2250\pm66)$ erg/cm$^3$, $M_0 = (1.615\pm0.001)$ $\mu_B$/Mn. The inset shows the geometry of FMR experiment: $n_U$ is the direction of easy axis of the uniaxial anisotropy.

Experiment with polarized neutrons was performed on a monochromatic reflectometer NREX (wavelength of 0.43 nm, the energy of 4.4 meV) located at the research reactor FRM II (Garching, Germany). The beam of polarized neutrons (polarization of 99.99%) fell on the sample at grazing angles $\theta = (0.15-1)$. The beam divergence $\Delta\theta_l = 0.025°$ was fixed by two apertures. The polarization of the reflected beam is detected using the analyzer with an efficiency of 98%. The external magnetic field in the experiment was directed in the sample plane and normal to the scattering plane. At a fixed temperature 4-channel intensity of small-angle reflection: $R^{++}(\theta)$, $R^{--}(\theta)$, $R^{+-}(\theta)$, $R^{-+}(\theta)$ were recorded. The + and - signs point the neutron spin projection on the external magnetic field. The reflection coefficients without neutron spin flip $R^{++}(\theta)$ and $R^{--}(\theta)$ (SF-scattering) are sensitive to the sum and difference of nuclear profile (SLD) and the magnetization lying in the plane of the substrate and collinear to the external field ($M_{\parallel}$) respectively. The reflection coefficients with neutron the spin-flip $R^{+-}(\theta)$, $R^{-+}(\theta)$ are sensitive to the component of the non-collinear to the external field magnetization $M_{\perp}$. A feature of reflectometry of polarized neutrons is the insensitivity to the component of the magnetization normal to the plane of the sample [31, 32]. Fig. 3a shows the reflection coefficients measured at $T > T_C$ for heterostructure #1. The reflection coefficients without spin flip (NSF-scattering) $R^{++}(\theta)$ and $R^{--}(\theta)$ are characterized by a region of total reflection at $Q < 0.15$ nm$^{-1}$ and Kiziha oscillations at $Q \approx Q_{crit}$ by the interference on different boundaries section in the structure. The essential difference between NSF -reflection coefficients $R^{++}(\theta)$, and $R^{--}(\theta)$, indicates the presence of collinear magnetization component. At the same time scattering with spin flip (SF- scattering) indicates on the presence of noncollinear magnetization component. Characteristic for SF- scattering of this structure is the existence of so-called the resonance peak near $Q_{crit}$, caused by resonance enhanced of neutron standing waves [31-33].

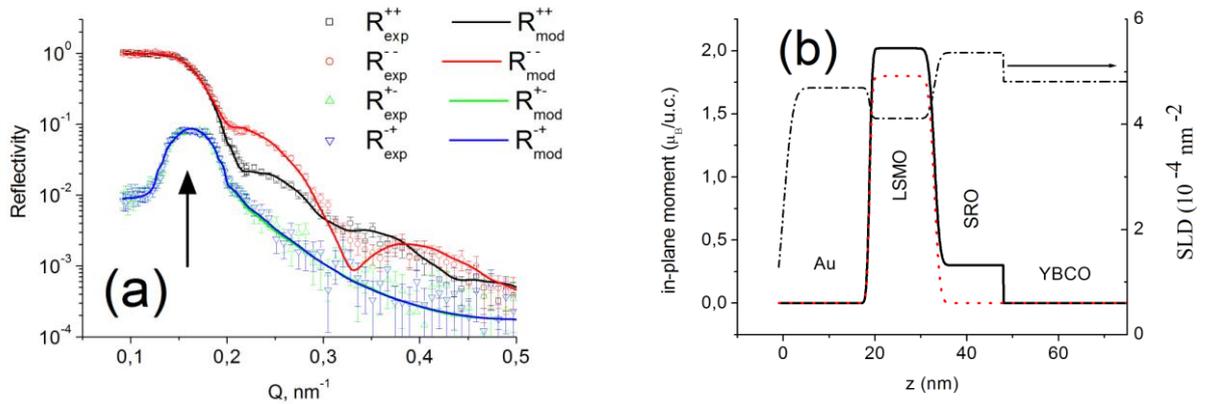

Fig. 3. (a) The reflection coefficients of neutrons from Au/LSMO/SRO/YBCO heterostructure #1 at $T = 80$ K, $H = 30$ Oe. The calculated curves of the reflection coefficients are shown by solid lines. The calculated curves for the spin-flip reflection coefficients $R^{+-}(\theta)$ and $R^{-+}(\theta)$ are coincided. The arrow indicates the position of the waveguide peak coinciding with the critical value of the transmitted torque total reflection $Q_{crit}$. (b) The nuclear profile SLD (dot-dash) and profiles of in plane magnetizations: $M_{\parallel}$ (solid line) and $M_{\perp}$ (dotted line) for the heterostructure.

The fit of the experimental data for the model proposed in [28] allowed to restore the profile of nuclear scattering length density (SLD), and the profile of both the collinear and non-collinear magnetic moment components in the plane (Fig. 3b). As follows from the fit, the interface has sharp transition region which does not exceed 2 nm. The magnetic state of the system is well described by the magnetic moment 2.5 $\mu_B$/Mn, at an angle of 43.3° to the direction of the external field. Projection of SRO film magnetization on the plane of substrate 0.3 $\mu_B$/Ru is parallel to the external magnetic field. The spin-flip scattering above $T_C$ is specified by the component of the magnetization LSMO film perpendicular directed to the external field.

3. Results and discussion

Table 1 shows the compositions and thicknesses of the films in heterostructures and the experimentally determined values of induced variation of the magnetic moment of the heterostructure – $\Delta m$. $\Delta m$ for samples #2 and #3 were obtained from FMR measurements. LSAT substrate was used for heterostructure #4, on which an epitaxial film of cuprate superconductor doped by Ca -YCBCO ($Y_{0.7}Ca_{0.3}Ba_2Cu_3O_x$) was deposited.

Table 1.

| N | Substrate | $d_S$, nm | $d_{SRO}$, nm | $d_{LSMO}$, nm | $\Delta m$, $10^{-6}$ emu |
|---|---|---|---|---|---|
| 1 | (001)LaAlO$_3$ | 80 | 20 | 14 | 10 |
| 2 | (110)NdGaO$_3$ | 80 | 17 | 7 | (5±1.5) |
| 3 | (110)NdGaO$_3$ | 180 | 0 | 20 | ≤(1±2) |
| 4 | (001)LSAT | 150 | 13 | 25 | 2.5 |
| 5 | (110)NdGaO$_3$ | 0 | 14 | 40 | - |
| 6 | (110)NdGaO$_3$ | 0 | 0 | 50 | - |

$d_S$, $d_{SRO}$, $d_{LSMO}$–are thicknesses of YBCO, SRO and LSMO film correspondingly; $\Delta m$ is the change of magnetic moment. The variations of $m$ for the whole heterostructure were measured for heterostructure #1 and #4; in case of heterostructures #2 and #3 the ferromagnetic part of the heterostructure was measured. Cuprate superconductor $Y_{0.7}Ca_{0.3}Ba_2Cu_3O_x$ was deposited for heterostructure #4. There is no superconductor in heterostructures #5 and #6.

3.1. Magnetic measurements

Figure 4 shows a family of temperature dependences of the magnetic moment for heterostructure #1 that parallel to the plane of the substrate $m_\parallel$ for several external magnetic fields. These data were obtained using a SQUID magnetometer under cooling in a magnetic field (FC mode). The external magnetic field is in the plane of the substrate and was directed along one of its edges. More detailed measurements of magnetic anisotropy showed that the edge of the substrate forms an angle of 40 - 50 degrees relative to the easy axis of the magnetic anisotropy of LSMO. In the temperature range $T_C<T<T_{SRO}$ ($T_C \approx 55$ K is superconducting transition temperature for YBCO film and $T_{SRO} \approx 150$ K is the Curie temperature for SRO film) the magnetic moment of the heterostructure is equal to the sum of magnetic moments of LSMO and SRO films. Under the influence of magnetic field the total magnetic moment changes due to the interaction between the magnetic moments of the films of LSMO and SRO. At low magnetic filed the total magnetic moment is smaller than the magnetic moment for LSMO film at the same temperature but at higher magnetic fields $H \geq 1$ kOe the total magnetic moment is larger than the one for the LSMO film.

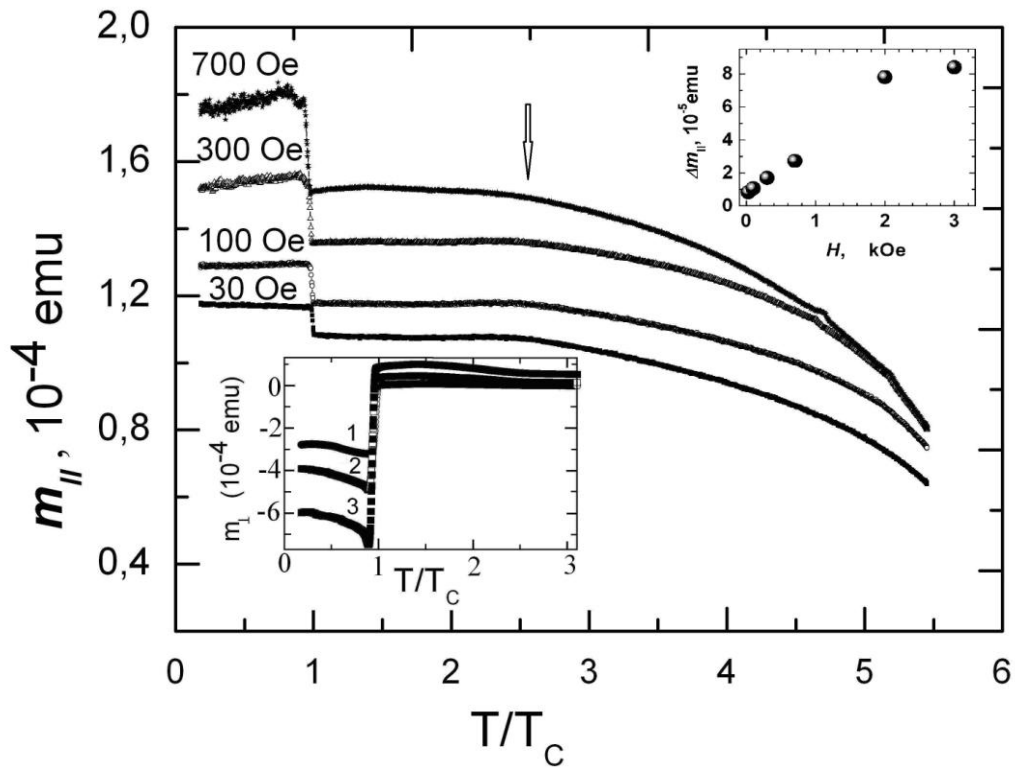

Fig. 4. Temperature dependence of the magnetic moment for $m_\parallel$ for heterostructure #1 cooled in magnetic fields (FC mode) of $H$ = 30, 100, 300, 700 Oe. The SRO film transition temperature to a ferromagnetic state is indicating by arrow. Left lower inset: Dependence of $m_\perp$ magnetic moments in a field perpendicular to the substrate plane, measured in magnetic fields 1 – 30, 2 – 200, 3 - 500 Oe. The top right inset shows the dependence of the magnetic moment $\Delta m_\parallel$ heterostructures on the external magnetic field at $T_C \approx T$.

At $T \approx T_C$ a sharp increase of the magnetic moment of the heterostructure was observed (Fig. 4). When the thickness of YBCO film $d_S = 80$ nm less than the London penetration depth the magnetic field which is directed along the plane of the film, fully penetrates into the superconductor. Diamagnetic response due to the Meissner effect is not observed. Note, if the sample plane is tilted towards a small angle relative to the direction of the external magnetic field (which is very difficult to avoid in the absence of special adjustment), the influence of the perpendicular magnetic field is substantially greater than the parallel due to the demagnetizing factor of the film [34]. But the influence of the perpendicular component of the magnetic field is substantially reduced in cooling in a magnetic field (FC regime) due to the occurrence of Abrikosov vortices [34]. Indeed, $\Delta m_\parallel$ (T) dependence changes in zero magnetic field magnetic fields cooling (ZFC-regime). Influence of the perpendicular component leads to the fact that the temperature dependence of the magnetic moment is similar to the case of a perpendicular orientation (see the lower inset in Fig. 4).

According to [35, 36] the induced magnetic moment in a superconductor can be estimated as $\Delta m_s \approx 2\mu_B N_S k_B T_C V \approx 10^{-5}$ emu where the density of states $N_S = 7 \cdot 10^{22}$ (eV·cm$^3$)$^{-1}$ [37], $k_B$ and $V$ are the Boltzmann constant, and volume of YBCO film correspondingly. This estimate is the same order of magnitude with the measured values $\Delta m_\parallel$ at $T \approx T_C$ (see. Fig. 4). According to calculations [7], ferromagnetic exchange field induces a magnetic moment in a superconductor at a distance from interface of the order of the coherence length. There is increasing of $\Delta m_\parallel$ at $T \approx T_C$ (positive sign). The calculations [35, 36] predicted negative sign of $\Delta m_\parallel$. However, in [7] is stated that if there is ferromagnetic with negative magnetic polarization, the sign of magnetic moment variation could be changed. A changing of sign of magnetic moment in superconductor can be caused by orbital reconstruction at the interface YBCO/SRO [23].

In the case of the orbital reconstruction mechanism it was obtained that in YBCO/LCMO superlattices the induced magnetic moments of the Cu atoms in a superconductor is equal to 0.23$\mu_B$/Cu and it is directed against Mn moments [17]. However, there is no indication for which temperature this value was obtained. At the same time, in [19] for the same superlattices, 0.013 $\mu_B$/Cu was obtained at $T = 28$ K. If we assume that variation of the magnetic moment of our heterostructure is due to copper atoms positioned in a layer thickness of about 10 nm, for observation the variation of magnetic moment for heterostructure #1($\Delta m$ ~ $10^{-5}$ emu) (see. fig. 4) we should take the value of the induced magnetic moment 0.25 $\mu_B$/Cu. Changing the direction of the magnetic moment of copper was observed in [23], and in this case may be due to negative magnetization of SRO film [7, 26].

The magnitude of $\Delta m_{\parallel}$ increases with the external magnetic field (upper inset in Fig. 4). The calculation of the magnetic moment of the superconductor, which is in contact with a magnetic spin valve shows that the induced magnetic moment is proportional to $\sin\theta$, where $\theta$ is misorientation angle in the ferromagnetic spin valve [35]. As follows from Fig. 4, at low fields the projection of the magnetic moment of the SRO (($m_{SRO}$) film on the plane of the substrate is directed opposite to the magnetic moment of the LSMO film ($m_{LSMO}$) that means that $\theta$ is close to 180°. Increasing the magnetic field $\theta$ reduces and $\Delta m_{\parallel}$ increases, which is observed in experiment. Depending on direction of $\theta$ changing with *H* (clockwise or *anti clockwise*), $\Delta m$ could be either positive or negative [35]. Our measurements show that at a field of about 1000 Oe, the SRO film contribution to the total magnetization of the heterostructure vanishes (temperature dependence of the heterostructure magnetic moment is close to $m_{LSMO}$ film). The angle between the $m_{SRO}$ and $m_{LSMO}$ is close to 90°. With further increase of *H*, increasing of $\Delta m_{\parallel}$ was observed although it should be saturated [35]

Change of $m_{\parallel}$ at $T \approx T_C$ can also be associated with the occurrence of anomalous magnetic response in the normal metal (Au), which is in contact with the superconducting spin valve S/F$_1$F$_2$, generating triplet superconducting correlation [11]. According to our estimation (using the value of the susceptibility $\chi = 10^{-3}$ [11]), this contribution does not exceed $2 \cdot 10^{-10}$ emu, which is significantly less than the measured values change $m_{\parallel}$.

If the magnetic field is perpendicular to the plane of the superconducting film the shielding currents flow in a layer at the edge of the film $\lambda_{\perp}=\lambda_L^2/d_S \approx 0.3$ μm. The magnetic field is expelled from the superconducting film and diamagnetic response (Meissner effect) is observed. It can be seen in the dependence of the magnetic moment of heterostructure, measured in the direction of the magnetic field directed perpendicular to the plane of the substrate $m_{\perp}(T)$ (see the lower inset in Fig. 4). The critical temperature of the superconductor T$_C$ in the heterostructure can be determined from $m_{\perp}(T)$. Note that $m_{\perp}(T)$ does not depend on the mode of cooling (FC or ZFC).

3.2. Ferromagnetic resonance

As noted earlier, the processing of the angular dependence of FMR spectra using (1) allows us determine the parameter $M_0$ and the direction of the easy axes in the LSMO film of the heterostructures. The Fig. 5 shows the temperature dependence of the parameter $M_0$ (left scale) as well as angles $\varphi_u$ and $\varphi_c$ under which easy axis of either uniaxial or biaxial magnetic anisotropy are directed correspondingly. The values of the angles are given for the case when the external magnetic field is directed along the [1-10] NGO substrate, which corresponds to the direction of the easy axis of the plane uniaxial anisotropy for LSMO film [25]. The Fig. 5 shows that the direction of easy axis does not change in the studied temperature range, and thus determined by the same reasons as at room temperature [25]. Here one must specify that the parameter $M_0$ determines the value of the magnetization in LSMO film ($M_{LSMO}$) only at temperatures above the transition to the ferromagnetic state for SRO film. At lower temperatures, should take into account inter-exchange between two ferromagnets [38] that would lead to a resonant relationship, different from formula (1).

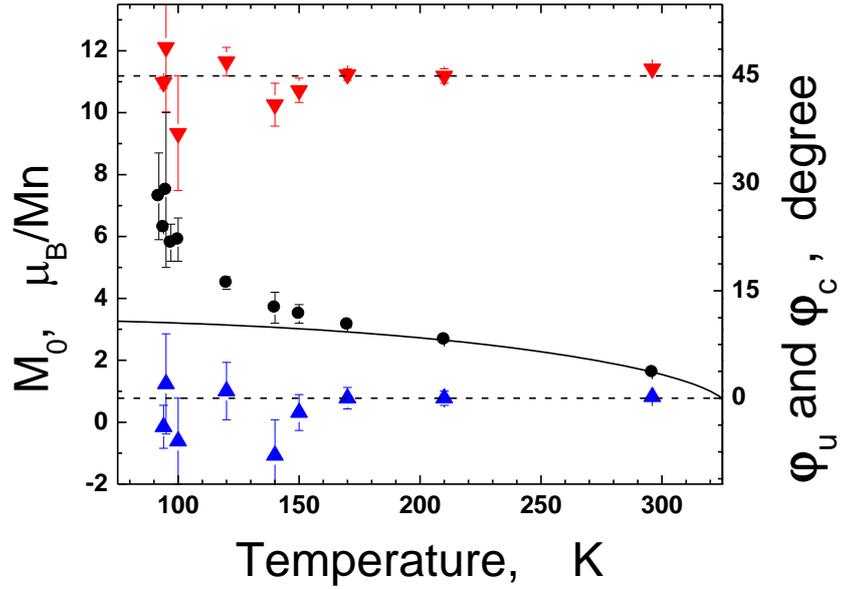

Fig.5. The temperature dependence of the parameter $M_0$ (circles, left scale) and direction of easy axis anisotropy (right scale) uniaxial ($\varphi_u$, triangles) and biaxial ($\varphi_c$, inverted triangles) for the LSMO layer heterostructure #2, obtained from the processing of the angular dependence of FMR spectra of using the formula (1). Solid line is calculated magnetization of LSMO film, and dashed lines demonstrate the absence of temperature dependence (see text).

To demonstrate this fact the calculated curve for magnetization layer LSMO, obtained with the use of molecular theory of Weiss [39] and the magnetization $M_{LSMO}$ defined by the eformula (1) in the temperature range above $T_{SRO}$ are shown on tha Fig. 5 by solid line. It can be seen from the Fig.5 that at $T > T_{SRO}$ the parameter $M_0$ exceeds the calculated curve that indicates on the presence of interlayer exchange interation between LSMO and SRO films. At $T \approx T_{SRO}$ magnetization of LSMO reaches 3.0 $\mu_B$/Mn for heterostructure #2 that agreed with the data for #1 obtained using a SQUID magnetometer (see. Fig. 4) if we take into account the ratio between the volumes of LSMO film in heterostructure and the unit cell for LSMO film.

Figure 6 shows the FMR spectra of LSMO film for heterostructure #2 near $T_C$. A huge non-resonant absorption signal at low magnetic fields having a hysteresis in the magnetic field is observed at $T \leq T_C$, when the YBCO film becomes superconducting [40]. As a result, the FMR signals at $T < T_C$ is recorded on the background of a giant non-resonant absorption. It increases the measurement error of the resonance field $H_{CF}$, but allows determining the temperature of the superconducting transition in the YBCO film $T_C$. At $T > T_C$, the value of $H_{CF}$ determined much more accurately, as shown in the insert in Fig. 6.

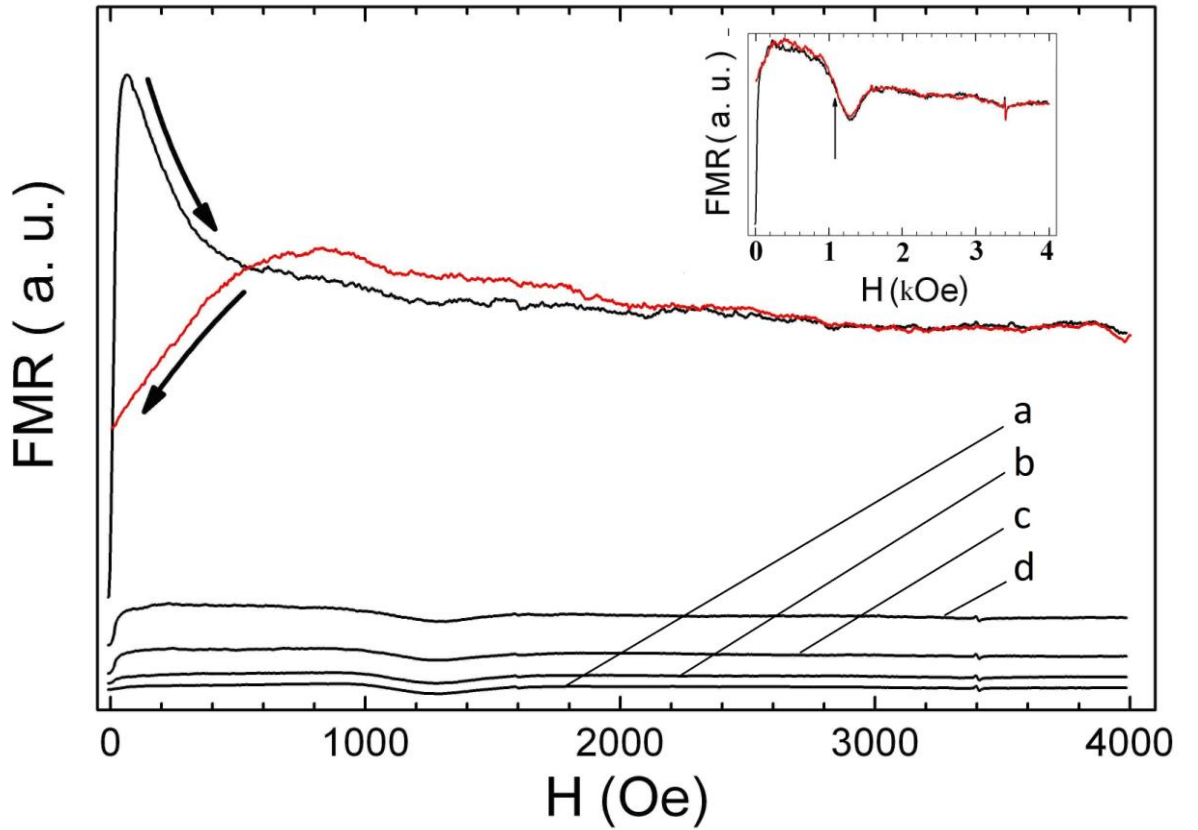

Fig.6. Temperature dependence of FMR spectra for heterostruture #2 at T on 1.04$T_C$ (a), 1.03$T_C$(b), 1.02$T_C$(c) and 1.01$T_C$ (d) above $T_C$ and spectrum at $T<T_C$. The sweep directions of the magnetic field at $T<T_C$ are indicated by arrows. FMR spectrum for two directions of magnetic filed variation for $T=1.01T_C$, is shown in inset. Arrow indicates the resonance field, $H_{CF}$.

Fig. 7 shows the temperature dependence of the resonance field $H_{CF}$ for FMR signals from the LSMO film in heterostructures #2 and #3 at $T \approx T_C$. In all cases, the direction of the external magnetic field is taken along the easy axis magnetization. We see that for heterostructures #2 there is a sharp change in the resonant field near $T_C$.

Previously, a reduction of the effective magnetization (an increased resonance field) of ferromagnetic layer in a bilayer S/F structure of the V/PdFe was observed in the spectrum of FMR in vicinity of $T_C$ in vanadium [12]. The effect was explained by the appearance of a cryptoferromagnetic state [41]. The authors of [7] suggested another explanation for the experimental data associated with a magnetic proximity effect and based on calculations using the model of quasi-classical approximation.

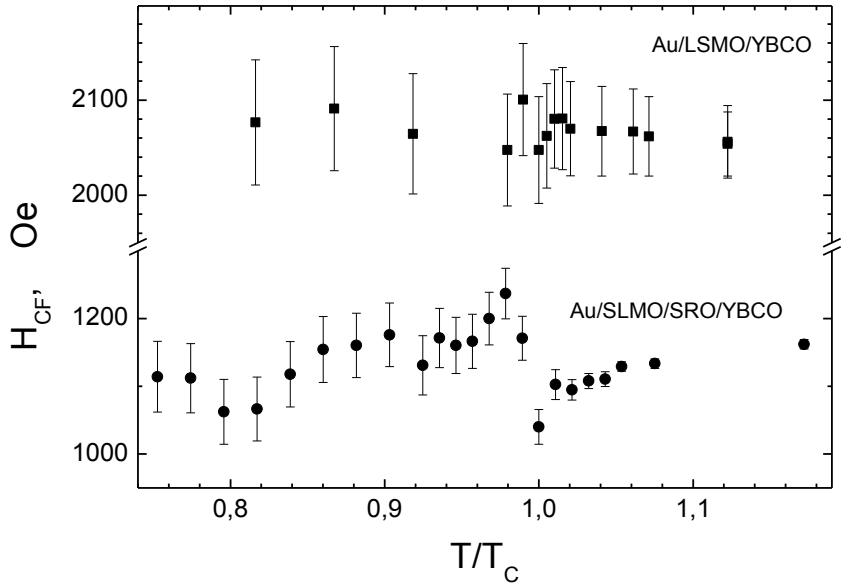

Fig. 7. Temperature dependence of the resonance field for heterostructures #2 (circles) and #3 (squares) in the vicinity of the superconducting transition temperature.

In the heterostructure #2 the LSMO film was separated from the superconducting YBCO film by a SRO ferromagnetic film. So the jump in the resonance field HCF in LSMO layer should be associated with a change in the magnetization in the SRO film. We must take into account the interlayer exchange interaction LSMO and SRO, which occurs through the magnetic ordering of the boundary layer with high conductivity [42-45]. The detailed calculation of interlayer exchange interaction in the structure of LSMO/SRO will be considered in a separate paper. Here we note that, using the procedure outlined in [38, 46], we got the relationship between frequency and resonance field for a layer of LSMO heterostructure LSMO/SRO similar to expression (1), but the value of the resonant field should be replaced by a combination:

$$H_{CF} + \frac{H_{J1}^{LSMO}\left(H_{CF} + H_{J1}^{SRO}\right)}{H_{SRO} - 4\pi M_{SRO} - H_{J2}^{SRO}}. \qquad (2)$$

Here $H_{SRO}$ and $M_{SRO}$ are the uniaxial magnetic anisotropy field and the magnetization of the film SRO correspondingly, $H_{J1}^{LSMO,SRO}$ and $H_{J2}^{SRO}$ effective field bilinear and biquadratic interlayer exchange for the corresponding layers, that are inversely proportional to the magnetization of the respective layers [38, 46].

Theresonant relations requires constant for combination (2) on both sides of the magnetization jump at $T \approx T_C$ because other parameters don't change within the range of about 1K. From the condition of the constancy we obtain the relationship between the resonance field jump $\delta H_{CF}$ in the LSMO film magnetization film and variation of $\delta M_{SRO}$ in SRO film:

$$\frac{\delta M_{SRO}}{M_{SRO}} \approx \frac{\delta H_{CF}}{H_{CF}} \frac{H_{SRO}}{4\pi M_{SRO}} \qquad (3)$$

Estimation by formula (3) shows that the change in the magnetization of the SRO film in the transition to the superconducting state of the YBCOfilm is about $0.5 M_{SRO}$. Taking into account the contribution of SRO film ($m_{SRO} \sim 10^{-5}$ emu) in the total magnetic moment $m_\parallel$ of the heterostructure (Fig. 4), the amount of change in the magnetic moment of a ferromagnetic composite is smaller than the magnetic moment induced in the superconductor. Note that a positive sign $\delta M_{SRO}$ indicates a reducing of the magnetization in the SRO film (see [7]).

Fig. 7 also shows that in the heterostructure #3 Au/LSMO/YBCO, where the YBCO film is in the contact with ferromagnetic LSMO film a noticeable change at $T \approx T_C$ is not detected within the measurement error. This difference in the heterostructures #2 and #3 can be explained by absence of the components of the triplet excitation of the superconducting correlation in ferromagnetic spin valve [1, 7, 27, 47, 48], and low transparency of YBCO/LSMO interface [47]. It leads to a negligible penetration of the superconducting order parameter of YBCO in the ferromagnetic spin-valve and therefore to negligible changes in the magnetic moment of the heterostructure #3.

Note that the value of the resonant field in the heterostructure #2 continues to reduce after the jump at $T \approx T_C$. We not attribute this behavior with not to increasing the magnetization which at these temperatures is near saturation but with increasing of the interlayer exchange interaction between LSMO and SRO films with decreasing temperature. The magnitude of the magnetization of LSMO at a temperature close to $T_C$ is 97% of the saturated value, which explicitly excludes its role in reducing of the resonant field with decreasing temperature. Furthermore, the temperature dependence of the magnetic anisotropy field also affects the value of $H_{CF}$.

As an example, Fig. 8 shows the temperature dependence of the resonance fields for LSMO film in various heterostructure. There is a wide variety of changes in $H_{CF}$ with decreasing temperature although the temperature dependence of the magnetization of LSMO in structures shown in Fig. 8 behave in the same way. It should be noted that for the heterostructures # 5 Au/LSMO/SRO/NGO Fig. 8 shows the appearance of an additional bending curve $H_{CF}(T)$ at $T \approx T_{SRO}$, indicating the existence of the interlayer exchange interaction between two ferromagnetic films. For heterostructure # 2 there is chang in $H_{CF}(T)$ at $T = T_C$ even on this scale

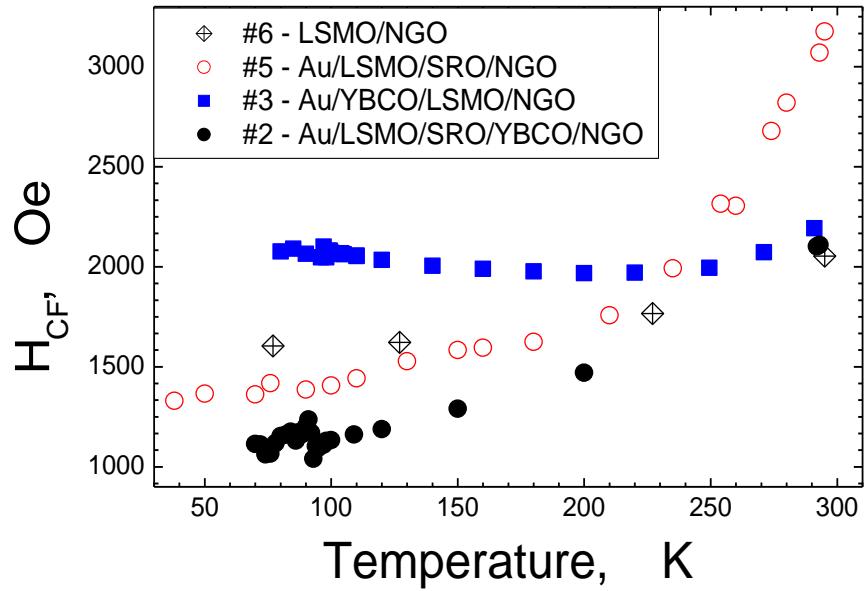

Fig. 8. The temperature dependences of the resonance field in LSMO films for heterostructures listed on the insert.

### 3.3. Neutron measurements

Neutron-reflection curves were measured for samples #1 and #4 in the temperature range $T =$ 10 - 80 K. For both samples an increase in spin-flip scattering (SF-scattering) at temperatures below $T_c$ was observed. Changing the magnetization of the heterostructure can be detected by means of the resonance peak caused by the neutron resonance enhanced standing wave. Fig. 9 shows the strength in SF-scattering near the resonance peak measured at temperatures both.

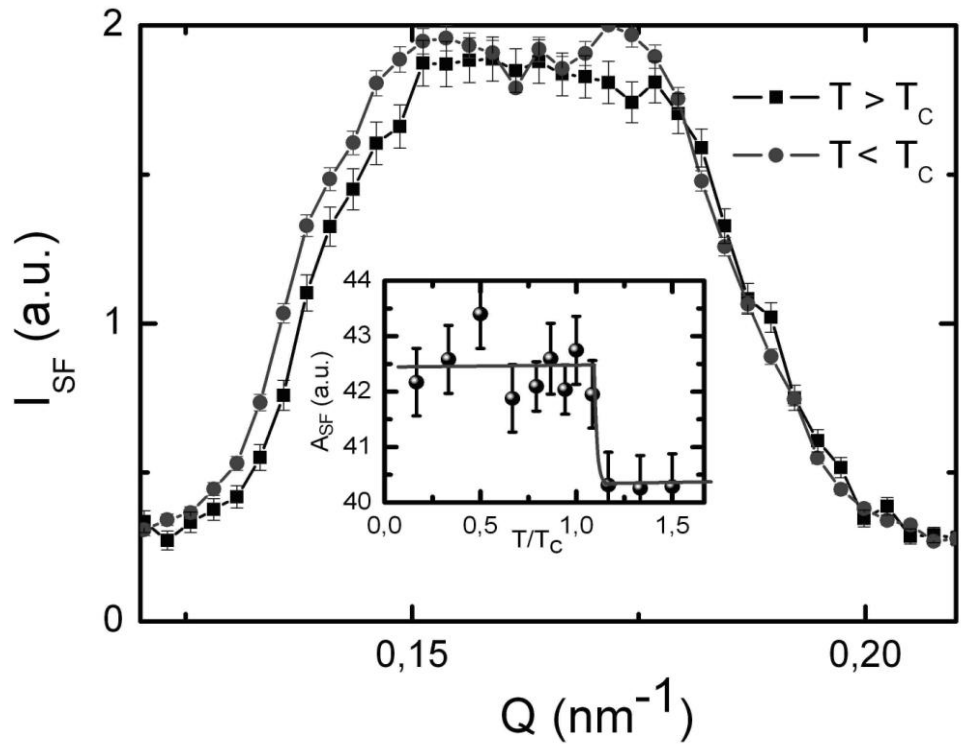

Fig. 9. The temperature dependence of the peak intensity of the neutron waveguide $I_{SF}(Q)$ for #1 heterostructure. The lines connect the data points. The inset shows the temperature variation of the peak area $A_{SF}$.

above and below $T_c$. Data is given for temperatures above and below $T_c$ summarized in the temperature range $T=70 - 90K$ and $T=10 - 60K$. As follows from Fig. 9, the the SF-scattering intensity increased during the transition to the superconducting state of the sample. The inset in Fig. 9 shows the temperature dependence of the peak intensity of the waveguide $I_{SF}(Q)$, which implies that an increase in SF-scattering at temperatures below $T_c$. A similar increase in SF-scattering intensity was also observed for sample #4 in either FC- or ZFC cooling.

Thus, the neutron data indicate a rise in the noncollinear magnetization in the heterostructure at the transition of YBCO film to superconducting state

For a heterostructure of #1 fitting the date shows that the growth of SF-scattering can be satisfactorily described by the appearance of magnetization in YBCO film with magnetization 0.4 $\mu_B$/Cu on a thickness 10 nm near interface with SRO. The vector of induced magnetization is directed parallel to the magnetization of LSMO layers, i.e on 45 ° to the external field. However, the change in the scattering without spin flip (NSF-scattering) (about 1-3%) caused by the moment does not exceed the statistical error in the measured reflection coefficients. The growth of SF-scattering at $T \approx T_C$ can be described by the appearance of a magnetic moment 4 $\mu_B$/Cu on the thickness on the order of the coherence length of YBCO (1 nm) near the interface with the SRO. This should result in a strong change of the coefficients NSF - scattering (over 10%) which was not observed in the experiment. Models involving growth noncollinear magnetization either in LSMO or SRO films describe the experimental neutron data also, but are in contradiction with the FMR data (see Section 3.2). A similar change in the peak intensity of a waveguide with a spin flip below the $T_C$ observed earlier in bilayer superconductor/ferromagnet - V(40nm)/Fe (1nm) [32, 49], which was interpreted as the appearance of induced magnetization +0.1$\mu_B$/V in 7 nm vanadium bordering with a thin layer of iron [30, 49].

4. Conclusions

We experimentally observed the appearance of the induced magnetic moment inside the superconductor in heterostructures based on a cuprate superconductor with oxide ferromagnetic spin valve. The magnitude of a magnetic moment in superconductor is close to calculated data both from the induced magnetic moment of the Cu atoms due to orbital reconstruction at the interface, and from the model that takes into account changes in the density of states in the superconductor at the interface of superconductor with ferromagnet. The characteristic length of penetration of the magnetic moment into superconductor significantly exceeds the coherence length in the cuprate superconductor, indicating the dominance of the induced magnetic moment of Cu.

The authors are grateful to V.A. Atsarkin, I.V. Borisenko, A.F. Volkov, A.V. Zaitsev, B. Keimer, Y.V. Kislinski, G. Logvenov, A.M. Petrzhik, A.V. Shadrin for help with the experiment and useful discussions. This work was supported by RFBR 14-02-00165, 14-07-

00258, 14-07-93105 and 14-22-01007, grant research school NSH-4871.2014.2 and equipment involving the The Swedish Research Infrastructure for Micro and Nano Fabrication (Myfab). Neutron measurements were carried out on the instrument NREX, the scientific community it served by Max Planck in the center of Heinz Mayer Lyaybnits (Heinz Maier-Leibnitz Zentrum), Garching, Germany

References

1. H. Zabel and S. D. Bader (Eds.), *Magnetic Heterostructures*, *Advances and Perspectives in Spinstructures and Spintransport*, Springer Tracts in Modern Physics, **227**, 251-289 (Springer, Berlin Heidelberg, 2008).

2. A. I. Buzdin Rev. of Mod. Phys. **77**, 935 (2005)

3. W.L. Lim, N. Ebrahim-Zadeh, J.C. Owens, et al., Appl. Phys. Let. **102**, 162404 (2013).

4. K. Halterman and O.T. Valls, Phys. Rev. B **66**, 224516 (2002).

5. R. Fazio and C. Lucheroni, Europhys. Lett. **45**, 707 (1999).

6. V.N. Krivoruchko, E.A. Koshina, Phys. Rev. B **66**, 0145621 (2002).

7. F.S. Bergeret, A.F. Volkov, and K.B. Efetov, Phys. Rev. B **69**, 174504 (2004).

8. F.S. Bergeret, A.L. Yeyati, and A. Martin-Rodero, Phys. Rev. B **72**, 064524 (2005).

9. M.Yu. Kharitonov, A.F. Volkov, and K.B. Efetov, Phys. Rev. B **73**, 054511 (2006).

10. R. Grein, T. Löfwander, and M. Eschrig, Phys. Rev **88**, 054502 (2013).

11. M. Alidoust, K. Halterman, J. Linder, Phys. Rev. B **89**, 054508 (2014).

12. I.A. Garifullin, D.A. Tikhonov, N.N. Garif'yanov, et al., Appl. Magn. Reson., **22**, 439 (2002).

13. M.G. Flokstra, S.J. Ray, S.J. Lister, et al., Phys. Rev. B. **89**, 054510 (2014).

14. J. Xia, V. Shelukhin, M. Karpovski, et al., Phys. Rev. Lett. **102**, 087004 (2009).

15. R.I. Salikhov, I.A. Garifullin, N.N. Garif'yanov, et al., Phys. Rev. Lett. **102**, 087003 (2009)

16. J. Stahn, J. Chakhalian, Ch. Niedermayer, et al., Phys. Rev. B **71**, 140509(R) (2005).

17. D.K. Satapathy, M.A. Uribe-Laverde, I. Marozau, et al., Phys. Rev. Lett. **108**, 197201 (2012)

18. J. Chakhalian, J. W. Freeland, G. Srajer, et al., Nat. Phys. **2**, 244 (2006)

19. J. Chakhalian, J. W. Freeland, H.-U. Habermeier, et al., Science **318**, 1114 (2007)

20. H -U Habermeier, Journal of Physics: Conference Series **108**, 012039 (2008)

21. J. Santamaria, J. Garcia-Barriocanal, Z. Sefrioul and C. Leon. International Journal of Modern Physics B Vol. **27**, No. 19 (2013)

22. J. Salafranca and S. Okamoto, Phys. Rev. Lett. **105**, 256804 (2010)

23. H.-U. Habermeier, G. Cristiani Physica C **408–410** 864 (2004)

24. E. Morenzoni, Th Prokscha, A. Hofer, et al., J. Appl. Phus., 81, 3341 (1997)

25. G. A. Ovsyannikov, A. M. Petrzhik, I. V. Borisenko, et al JEPT, **108,** 48 (2009))

26. G. Koster, L. Klein, W. Siemons, et al., Rev. Mod. Phys. **84**, 253 (2012).

27. https://www.qdusa.com/products/mpms3.html

28. Yu.N. Khaydukov, G.A. Ovsyannikov, A.E. Sheyerman, et al., Phys. Rev. B **90**, 035130 (2014)

29. G.A. Ovsyannikov, A.E. Sheyerman, A. V. Shadrin, et al., JETP Lett., 97, 145 (2013).

30. V.V. Demidov, I.V. Borisenko, A.A. Klimov et al.., JEPT, **112**, 825 (2011)

31. L.G. Prat, Phys.Rev., 95, 359, (1954)

32. Yu.N. Khaydukov, B. Nagy, J.-H.Kim, et al., Письма в ЖЭТФ, **98**, 116, (2013)

33. H. Zabel, K. Theis Bröhl, B. P. Toperver, in Handbook of Magnetism and Advanced Magnetic Materials, John Wiley & Sons 2007.

34. C. Monton, F. de la Cruz, and J. Guimpel. Phys. Rev. B**75**, 064508 (2007)

35. N.G. Pugach and A.I. Buzdin, Appl. Phys. Lett. **101**, 242602 (2012).

36. T. Lofwander, T. Champel, J. Durst, M. Eschrig Phys. Rev Lett., **95**, 187003 (2005).

37. J. Mannhart, Superconducting Science and Technology, 9, 49 (1996)

38. N.M. Kreines, Low Temperature Physics, **28**, 581 (2002)

39. C Kittel, "Introduction to Solid State Theory", Wiley, New York (1971).

40. V.A. Atsarkin, V.V. Demidov, N.E. Noginova, Superconductivity: Physics, Chemistry, Technology, **5**, 305 (1992) (in Russian).

41. F.S. Bergeret, K.B. Efetov, A.I. Larkin, Phys. Rev. B **62**, 11872 (2000)

42. P. Padhan, W. Prellier, and R. C. Budhani, Appl. Phys. Lett. **88**, 192509 (2006).

43. M. Ziese, I. Vrejoiu, E. Pippel, et al., Phys. Rev. Lett. **104**, 167203 (2010).

44. A.Y. Borisevich, A.R. Lupini, J. He, et al., Phys. Rev. B **86**, 140102(R) (2012).

45. M. Ziese, F. Bern, A. Setzer, et al., Eur. Phys. J. B **86**, 42 (2013).

46. V.N. Men'shov, V.V. Tugushev, JEPT, **98**, 123 (2004).

47. A.E. Sheyerman, K.Y. Constantinian, G.A. Ovsyannikov, et al, JEPT, **120**, 1024 (2015)

48. C. Richard, M. Houzet, J.S. Meyer, Phys. Rev. Lett. **110**, 217004 (2013)

49. Yu.N. Khaydukov, V.L. Aksenov, Yu.V. Nikitenko, et al., J. of Supercond. Nov. Magn. **24**, 961 (2011)